\documentstyle[pra,aps,twocolumn,epsfig]{revtex}
\input epsf
\draft
\begin{document}
\title{Minimum-error discrimination between three mirror-symmetric 
states}
\author{Erika Andersson$^1$, Stephen M. Barnett$^1$, Claire R. Gilson$^2$, 
and Kieran Hunter$^1$}
\address{$^1$Department of Physics and Applied Physics, University of 
Strathclyde, Glasgow G4 0NG, Scotland}
\address{$^2$ Department of Mathematics, University of Glasgow, Glasgow 
G12 8QQ, Scotland}
\date{Dec 14, 2001}

\maketitle
\begin{abstract} 
We present the optimal measurement strategy for distinguishing between three 
quantum states exhibiting a mirror symmetry. The three states live in a
two-dimensional Hilbert space, and are thus overcomplete. By mirror symmetry
we understand that the tranformation $\{|+\rangle\rightarrow |+\rangle ,
|-\rangle\rightarrow -|-\rangle\}$ leaves the set of states invariant.
The obtained measurement strategy minimises the error probability.
An experimental realization for polarized photons, realizable with current 
technology,
is suggested.
\end{abstract}


Quantum communication theory is concerned with the transmission of
information using quantum states and channels.
A sender encodes a message onto a set of signal states, and the task
for the receiving party is to decode the message as well as possible.
Suppose $|\psi_i\rangle$ is a set of $M$ quantum states, each 
occurring with probability $p_i$. This set, the ``letters of the alphabet'', 
and the prior probabilities are known also to the receiving party.

A general measurement strategy can be described in terms of a probability
operator measure (POM), also called a probability operator-valued 
measure \cite{hel,kraus,per}. 
The different measurement outcomes, labeled
by $j$, are associated with operators $\hat{\pi}_j$, called the elements 
of the POM. Given that the system to be measured is prepared
in state $|\psi_i\rangle$, the probability to obtain result $j$ is
\begin{equation}
p(j|i)=\langle\psi_i|\hat{\pi}_j|\psi_i\rangle.
\end{equation}
For a von Neumann measurement, the operators $\hat{\pi}_j$ are
projectors onto the orthonormal eigenstates of the observable to
be measured. In general, however, $\hat{\pi}_j$ are not orthogonal.
Nevertheless, all the eigenvalues of $\hat{\pi}_j$ have to be positive
or zero, and the POM elements sum to the identity operator, 
$\sum_j\hat{\pi}_j=\hat{\bf 1}$. These conditions reflect the facts
that probabilities are non-negative and that the measurement always
yields a result, even if the result might not provide any information.

The task is now to find an optimal measurement strategy based on the 
knowledge of the signal states and their prior probabilities.
One possibility is to choose the measurement strategy which minimises the
probability of error in assigning the correct signal state. 
If the prior probability for signal state 
$|\psi_i\rangle$ is $p_i$, the error probability is given by
\begin{equation}
P_{error}=1-\sum_{j=1}^M \langle\psi_j|\hat{\pi}_j|\psi_j\rangle p_j.
\end{equation}

The conditions which the minimum-error strategy must satisfy are 
known to be\cite{hel}
\begin{eqnarray}\label{pomcond1}
\hat{\pi}_j(p_j|\psi_j\rangle\langle\psi_j|-
            p_k|\psi_k\rangle\langle\psi_k|)\hat{\pi}_k&=&0 \quad 
\forall\quad j,k,\\
\label{pomcond2}
\sum_{j=1}^M p_j|\psi_j\rangle\langle\psi_j|\hat{\pi}_j-
             p_k|\psi_k\rangle\langle\psi_k|&\geq & 0 \quad \forall\quad k,
\end{eqnarray}
where the inequality in the second condition states that all the
eigenvalues of the operator on the left-hand side must be greater than
or equal to zero. These conditions are in general not transparent 
enough to be used for obtaining
the optimal solution. In fact, the optimal strategy is only known for
some special cases, including the case with only two  signal states 
\cite{hel}, symmetric states \cite{hel,ban,steve}, and equiprobable
states that are complete in the sense that a weighted sum of projectors
onto the states equals the identity operator \cite{yuen}. For linearly
independent states, the optimal measurement is known in some cases 
\cite{sas1}.


We will consider the situation in which the signal states are given by
\begin{eqnarray}
|\psi_1\rangle &=&\cos\theta |+\rangle +\sin\theta |-\rangle \nonumber\\
|\psi_2\rangle &=&\cos\theta |+\rangle -\sin\theta |-\rangle\\
|\psi_3\rangle &=&|+\rangle\nonumber,
\end{eqnarray}
with prior probabilitites $p_{1,2}=p$ and $p_3=1-2p$, where 
$0\le p \le 1/2$. Due to symmetry it is enough to consider
the range $0\leq \theta \leq \pi/2$. The states $|+\rangle$ and 
$|-\rangle$ are orthonormal basis states. As a special case, the 
so-called trine states are obtained when $\theta=\pi/3$ and all 
the probabilities are equal to $1/3$ \cite{trineexp}.

The transformation $\{|+\rangle \rightarrow |+\rangle, 
|-\rangle \rightarrow -|-\rangle\}$ leaves the set of signal states
unchanged.
We expect that the POM elements will exhibit the same symmetry
as the set of signal states, and are led to the Ansatz

\begin{eqnarray}
\label{eans}
\hat{\pi}_{1,2}&=&|\phi_{1,2}\rangle\langle\phi_{1,2}|,\nonumber\\
\hat{\pi}_3 &=& (1-a^2)|+\rangle\langle +|=|\phi_3\rangle\langle\phi_3|,
\end{eqnarray}
where 
\begin{equation}
|\phi_{1,2}\rangle=1/\sqrt{2}(a|+\rangle \pm |-\rangle),
\end{equation} 
with ``+'' referring to $|\phi_1\rangle$, ``-'' to $|\phi_2\rangle$,
and $0\le a \le 1$.

If $p$ is large enough (certainly if $p=1/2$), we might expect 
that the optimal measurement strategy is the one which distinguishes
optimally between $|\psi_1\rangle$ and $|\psi_2\rangle$. This
is the case when $a=1$ and
\begin{eqnarray}
\label{epom1}
\hat{\pi}_1&=&{1\over 2}(|+\rangle +|-\rangle )(\langle +|+\langle -|),
\nonumber\\
\hat{\pi}_2&=&{1\over 2}(|+\rangle -|-\rangle )(\langle +|-\langle -|),\\
\hat{\pi}_3&=&0.\nonumber
\end{eqnarray}

It is convenient to start by investigating when this strategy is optimal. 
The POM elements (\ref{epom1}) can be checked to satisfy condition 
(\ref{pomcond1}). The inequality condition (\ref{pomcond2}) 
is seen to hold for $k=1,2$. For $k=3$, using matrix notation, 
\begin{equation}
|+\rangle\equiv\left(\begin{array}{c}1\\0\end{array}\right);
|-\rangle\equiv\left(\begin{array}{c}0\\1\end{array}\right),
\end{equation}
it can be written
\begin{equation}
p(\cos\theta +\sin\theta)
\left(\begin{array}{c c} \cos\theta & 0\\ 0 & \sin\theta \end{array}\right)
-(1-2p)\left(\begin{array}{c c} 1 & 0\\ 0 & 0 \end{array}\right) \geq 0,
\end{equation}
which is satisfied if
\begin{equation}
\label{epcond}
p\geq {1\over {2+\cos\theta (\cos\theta +\sin\theta )}}.
\end{equation}
Hence the measurement strategy given by Eq. (\ref{epom1}), 
which distinguishes optimally between $|\psi_1\rangle$ and 
$|\psi_2\rangle$, is the best choice provided $p_{1,2}=p$ is large 
enough.

If $p$ is too small to satisfy the inequality (\ref{epcond}), 
we need to consider a three-element POM, as given by the original 
Ansatz in Eq. (\ref{eans}). The equality condition (\ref{pomcond1}) 
holds trivially for $j=1$ and $k=2$.
For $j=1,2$ and $k=3$ it leads to the requirement
\begin{equation}
\label{eacond}
a={{p\cos\theta\sin\theta}\over{1-p(2+\cos^2\theta)}}.
\end{equation}

With this value of $a$, the inequality condition (\ref{pomcond2}) 
for $k=3$ is satisfied, the smaller eigenvalue being zero.
To test the inequality condition for
$k=1,2$, it is convenient to make use of the fact that a $2\times 2$ 
Hermitian matrix with elements $a_{11}, a_{12}, a_{21}$ and $ a_{22}$
is positive semidefinite if and only if $a_{11}, a_{22}$ and the 
determinant of the matrix are all greater that or equal to zero.
Upon evaluating the l.h.s. of condition (\ref{pomcond2}) in matrix form,
one finds that its elements $a_{11}$, $a_{22}$ are both greater than zero. 
The determinant turns out to be zero when the required value of $a$ is used. 

\begin{figure}
\centerline{\epsfxsize=7cm \epsffile{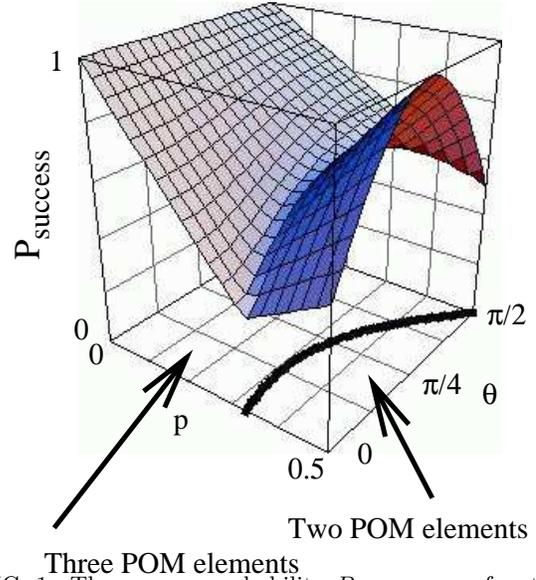}}
\caption{The success probability $P_{success}$ as a function of $p$ and 
$\theta$. The optimal measurement strategy will have either two or three
elements.}
\label{fregimes}
\end{figure}

It follows that this is indeed the measurement strategy which 
minimises the error probability. As shown in Fig. \ref{fregimes},
there are two regimes depending on
$p$ and $\theta$. If $p_{1,2}=p$ is large enough to satisfy
the inequality (\ref{epcond}), the strategy with $\hat{\pi}_3=0$, 
which distinguishes optimally between $|\psi_1\rangle$ and 
$|\psi_2\rangle$, is the solution. If $p$ is smaller than this, the full
three-element measurement is needed, with $a$ given by Eq. (\ref{eacond}).
The crossover between these two regimes corresponds to 
$p=(2+\cos\theta (\cos\theta +\sin\theta ))^{-1}$, where the
two strategies will coincide. 

It is straightforward to calculate the success probability, that is,
the probability of correctly identifying the signal state, of the
devised optimal measurement strategy.
For $p\geq 1/\left[2+\cos\theta (\cos\theta +\sin\theta )\right]$,
we find 
\begin{equation}
P_{success}=1-P_{error}=p(1+\sin 2\theta).
\end{equation}
As a special case, for $p=1/2$ with $0\le\theta\le\pi /2$, this 
reduces to the probability to distinguish with minimal error
between two non-orthogonal states.
For $p=1/2$ and $\theta =\pi /4$, the success probability is equal 
to one, corresponding to the situation where the two states are 
orthogonal.

For $p\leq 1/\left[2+\cos\theta (\cos\theta +\sin\theta )\right]$,
the success probability is
\begin{equation}
P_{success}={{(1-2p)\left[ p\sin^2\theta + 1-2p-p\cos^2\theta \right]}
\over{1-2p-p\cos^2\theta }},
\end{equation}
which reduces to 2/3 for equal probabilities $(p_{1,2}=p_3=1/3)$,
as long as $\pi /4\le\theta\le\pi /2$.

In many of the measurement situations where the optimal measurement
strategy is known, the optimal strategy is the
square-root or ``pretty good'' measurement \cite{ban,sas1,haus}
with POM elements
\begin{equation}
\pi_i=\hat{\rho}^{-1/2}p_i|\Psi_i\rangle\langle\Psi_i|\hat{\rho}^{-1/2},
\end{equation}
where
\begin{equation}
\hat{\rho}=\sum_{i=1}^3p_i|\Psi_i\rangle\langle\Psi_i|.
\end{equation} 
It is interesting to note that the square root measurement strategy
will coincide with the optimal strategy for some nontrivial $p$ and 
$\theta$, but is in general not the optimal strategy.


In a way similar to recent optical realisations of generalised
measurements \cite{trineexp,huttner,helexp,unaexp}, it is possible 
to realise the devised measurement strategy for photon polarization. 
The states 
$|+\rangle$ and $|-\rangle$ will now correspond to orthogonal 
polarisation states of a single photon, for example horizontal 
and vertical linear polarisation $|H\rangle$ and $|V\rangle$.
The realisation relies on the fact that any generalised measurement
can be extended to a projective (von Neumann) measurement in a 
higher-dimensional Hilbert space, the so-called Naimark
extension \cite{hel,kraus,per,nai}. Physically, this can be achieved
by coupling the system to an ancilla, or for example as in the optical 
realisations, by the introduction of additional input ports to the system, 
thus extending the Hilbert space.

\begin{figure}
\centerline{\epsfxsize=5cm \epsffile{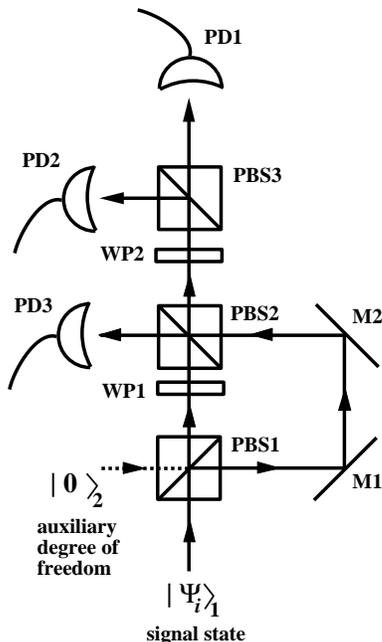}}
\caption{Optical network for implementing the minimum-error measurement 
on the three mirror-symmetric states. The state to be measured is
incident on PBS1. PBS1, PBS2 and PBS3 are polarising beam splitters, 
transmitting horisontally polarised light and reflecting vertically
polarised light.
WP1 and WP2 are waveplates, and M1 and M2 are mirrors.}
\label{fnetwork}
\end{figure}

An optical network with beam splitters and wave plates can be used to 
couple the different polarisation states, and the measurement
result is obtained by detecting where the photon exits, as shown
in Fig. \ref{fnetwork}.
The signal state is incident on one of the input ports of the 
polarising beam splitter PBS1. In the case we are considering,
one auxiliary degree of freedom is needed, and this is provided by 
the second input port of the beam splitter. Only vacuum is
incident through this input. 

In matrix notation, for correct settings of the waveplates \cite{wave},
the optical network effects the unitary transformation
\begin{equation}\label{eumatrix}
U={1\over\sqrt{2}}\left(\begin{array}{c c c}
a & 1 & \sqrt{1-a^2}\\
a & -1 & \sqrt{1-a^2}\\
\sqrt{2(1-a^2)} & 0 & -\sqrt{2}a
\end{array}\right),
\end{equation}
using
\begin{equation}
|H\rangle_1\equiv\left(\begin{array}{c}1\\0\\0\end{array}\right),
|V\rangle_1\equiv\left(\begin{array}{c}0\\1\\0\end{array}\right),
|V\rangle_2\equiv\left(\begin{array}{c}0\\0\\1\end{array}\right).
\end{equation}
Here $|H\rangle_1$ and $|V\rangle_1$ are used to encode the signal state,
$|V\rangle_2$ being an auxiliary degree of freedom provided by the
second input port of PBS1. The mode $|H\rangle_2$ is not coupled
to the signal state modes, as horizontally polarized light incident
through the second input port of PBS1 will be transmitted by both beam 
splitters, always resulting in detection at PD3.
The rows of the matrix $U$ are simply the POM
element vectors $|\phi_{1,2,3}\rangle$, but extended in the third
auxiliary dimension so that the matrix row vectors are orthogonal.

The measurement at the output then distinguishes between horisontally 
and vertically polarised photons in mode 1 (PD1 and PD2, outcome 1 and 2), 
and (vertically polarised) photons in mode 2 (PD3, outcome 3). If the 
incident state is $|\psi_i\rangle$, the state after the network is 
\begin{equation}
U|\psi_i\rangle=\sum_{j=1}^3|j\rangle\langle\phi_j|\psi_i\rangle,
\end{equation}
where $|1\rangle\equiv |H\rangle_1$, $|2\rangle\equiv|V\rangle_1$
and $|3\rangle\equiv |V\rangle_2$.
It is easy to confirm that
\begin{eqnarray}
p(j|i)&=& \langle\psi_i|\hat{\pi}_j|\psi_i\rangle\nonumber\\
&=&\langle j|U|\psi_i\rangle\langle\psi_j|U^\dagger|j\rangle,
\end{eqnarray}
so that the setup indeed realises the intended measurement strategy.

We want to stress that an optical setup like this is straightforward
to realise. Indeed, a slightly different, but essentially equivalent
network was used in the experiments of Clarke {\it et al.} 
\cite{trineexp,unaexp}. Another equivalent network was suggested, for 
a different measurement task, by Sasaki {\it et al.} \cite{sas}, and
has recently been realized experimentally \cite{mizuno}.
In summary, we have obtained the measurement strategy which minimises 
the error probability when distinguishing between three mirror-symmetric
states. This example is noteworthy in that the number of non-zero POM
elements, for the minimum-error strategy, depends on the parameters
chosen for the set of states. It will be interesting to see if this 
unusual property persists for other measures of detection strategy
such as mutual information \cite{sas} and fidelity \cite{fidelity}.



We would like to thank P. \"Ohberg for his assistance in preparing the
manuscript.
This work was supported by the European Commission Marie Curie 
Fellowship scheme, the Royal Society of Edinburgh, the Scottish Executive
Education and Lifelong Learning Department, the International Centre
for Theoretical Physics and the UK Engineering and Physical Sciences
Research Council.


\end{document}